\documentstyle[12pt,epsf]{article}
\newcommand{\be}{\begin{equation}}
\newcommand{\bea}{\begin{eqnarray}}
\newcommand{\eea}{\end{eqnarray}}
\newcommand{\beas}{\begin{eqnarray*}}
\newcommand{\eeas}{\end{eqnarray*}}
\newcommand{\ba}{\begin{array}}
\newcommand{\ea}{\end{array}}
\newcommand{\ee}{\end{equation}}

\newcommand{\tr}{{\rm Tr}\ }
\def\identity{{\rlap{1} \hskip 1.6pt \hbox{1}}}
\newcommand{\nbox}{{\,\lower0.9pt\vbox{\hrule \hbox{\vrule height 0.2 cm \hskip
0.2 cm \vrule height 0.2 cm}\hrule}\,}}

\newcommand{\pa}{\partial}

\newcommand\text[1]{\rm #1}

\newcommand{\CC}{\hbox{\xiiss C\kern-.4emI}}
\newcommand{\RR}{\hbox{\xiiss R\kern-.45emI}}
\newcommand{\ZZ}{\hbox{\xiiss Z\kern-.4emZ}}
\newcommand{\CCs}{\hbox{\ixss C\kern-.4emI}}
\newcommand{\ZZs}{\hbox{\ixss Z\kern-.4emZ}}
\newcommand{\pasl}{\pa\kern-.55em /}
\def\href#1#2{#2}

\begin{document}
\begin{titlepage}
\hfill
\vbox{
    \halign{#\hfil         \cr
           hep-th/0403289 \cr
           } 
      }  
\vspace*{20mm}
\begin{center}
{\Large \bf Generally Covariant Actions \\ for Multiple D-branes}

\vspace*{15mm}
\vspace*{1mm}
{Dominic Brecher, Kazuyuki Furuuchi, Henry Ling, Mark Van Raamsdonk} 

\vspace*{1cm}

{Department of Physics and Astronomy,
University of British Columbia\\
6224 Agricultural Road, 
Vancouver, B.C., V6T 1W9, Canada}

\vspace*{1cm}
\end{center}

\begin{abstract}

We develop a formalism that allows us to write actions for multiple D-branes with manifest general covariance. While the matrix coordinates of the D-branes have a complicated transformation law under coordinate transformations, we find that these may be promoted to (redundant) matrix fields on the transverse space with a simple covariant transformation law. Using these fields, we define a covariant distribution function (a matrix generalization of the delta function which describes the location of a single brane). The final actions take the form of an integral over the curved space of a scalar single-trace action built from the covariant matrix fields, tensors involving the metric, and the covariant distribution function. For diagonal matrices, the integral localizes to the positions of the individual branes, giving $N$ copies of the single-brane action.
\end{abstract}

\end{titlepage}

\vskip 1cm
\section{Introduction}

There is probably no endeavor in the realm of string theory that has led to more revelations than contemplating collections of D-branes \cite{polchinski}. The list is impressive: the microscopic derivation of black hole entropy \cite{sv}, Matrix theory \cite{bfss}, gravity/gauge theory duality \cite{maldacena}, noncommutative geometry in string theory \cite{cds, dh, sw}, and so forth. Further, collections of D-branes might well play a role in real-world phenomenology if it turns out that the standard model fields arise from modes on D-branes or that D-branes play a role in cosmology either as topological defects, as part of a stringy mechanism for inflation, or as a novel type of matter in the early universe.\footnote{The references for applications of D-branes to phenomenology or cosmology are too numerous to list here.}

Despite all of this, our understanding of D-brane physics is far from
complete. At the most basic level, one would like to know the actions
that govern collections of D-branes in general situations. At the very
least, these actions should be consistent with the principles and
symmetries upon which string theory is based. Yet, so far, we have
failed to incorporate one of the most important of all of these,
general covariance, into the actions for more than a single D-brane
(though significant progress has been made
\cite{dbs,dbs2,hassan}; see also \cite{douglas2,douglas,douglas3} for important earlier work.)

The basic difficulty is well known. When two or more D-branes nearly
coincide, the number of light degrees of freedom goes like the square
of the number of branes since these degrees of freedom arise from open
strings that begin on one brane and end on another. For example, with $N$ D0-branes (or the transverse coordinates of higher-dimensional branes) the configurations are described by $N \times N$ Hermitian matrices $X^i$ rather than by collections of points \cite{witten}. When all the matrices commute, their simultaneous eigenvalues $x^i_n$ correspond to well defined locations for the $N$ branes. But for general noncommuting configurations, the branes do not have well defined locations and the geometrical picture is some ``fuzzy'' higher-dimensional object described by noncommutative geometry. Since it is difficult to say exactly where the branes are in this case, it is not at all clear how to implement a local spacetime symmetry such as general covariance. 

The problem boils down to two questions: 
\begin{itemize}
\item ``How do the matrix coordinates of D-branes transform under a change of spacetime coordinates?'' 
\item ``What actions are invariant under these transformations?'' 
\end{itemize}
The goal of this paper is to provide answers to both of these (though our answer to the first question will not be completely explicit). For the most part, we consider D-particles and restrict to spatial diffeomorphisms, since many of the interesting issues arise already in this simple case. 

The first question amounts to asking how the transformation law $x^i \to F^i(x)$ generalizes when $x^i$ is a matrix. In section 2, we outline various consistency conditions that such a generalization must satisfy, for example, that the result should respect the multiplication (composition) law of the diffeomorphism group. We find that a generalization consistent with all of our requirements appears to be possible, however the result is very complicated, and (as previously noted in \cite{dbs}) necessarily involves the metric connection and its derivatives in addition to the transformation function $F$.

In section 3, we propose a more radical approach. We note that for any point $x$ on a manifold, there is a naturally associated vector field $v^i_y$ in the neighborhood of the point, defined such that $v^i_y$ points in the direction of the geodesic from $y$ to $x$ and has length given by the geodesic distance. This vector field is well defined globally if the manifold is such that there is a unique geodesic between any pair of points. In this case (to which we restrict), we can describe the location of a D-brane by a vector field (whose exponential map is constant) rather than by a spacetime point. While specifying a spacetime point would seem to be a much more efficient description in the case of a single D-brane, the covariant transformation law for the vector field, $v^i_y \to \partial_j F^i(y) v^i_y$ is much simpler (trivial, in fact) to generalize to the case when $v^i_y$ is a matrix. Thus, we are motivated to look for some matrix valued object $V^i_y$ built from $X$ and the metric that transforms as a vector field under diffeomorphisms.

It turns out that such an object is guaranteed to exist as long as there is a consistent transformation rule.
That is, given a transformation rule for $X^i$ satisfying the requirements of section 2, we can directly and explicitly construct an object $V_y(X,g)$ which transforms covariantly as a vector field. Further, we describe an algorithm to determine the relationship between $X$ and $V_y$ without reference to the transformation law. 

Armed with the covariant object $V_y$, we proceed in section 4 to construct generally covariant actions. These take the form of an integral over the transverse space of a scalar Lagrangian density constructed from $V_y$ and tensorial quantities built from the metric. Since the D-branes are objects with some extent in the transverse directions when the matrix coordinates do not commute, it is quite natural that the actions should take the form of an integral rather than an expansion about some fixed point. However, in order that the action be localized to the vicinity of the branes, we are led to introduce one additional ingredient: a covariant matrix distribution function $\delta(V_y)$ which reduces to a collection of delta functions for diagonal matrices, but to some smooth extended distribution for general noncommuting matrices. The final actions take the form 
\be
\label{intact}
S = \int dt \int d^d y \; \sqrt{g(y)} \; {\rm Tr}({\cal L}(V_y, g(y), R(y), \nabla R(y) \dots) \delta(V_y)) \; .
\ee
Since ${\cal L}$ is an arbitrary scalar, there are many actions consistent with general covariance; in fact we show that all generally covariant multiple D-particle actions can be written in this way by describing a minimal basis of invariant actions, each of the form (\ref{intact}).

Our results are completely consistent with previous work of de Boer and Schalm. In \cite{dbs}, these authors proposed a method, ``base-point independence'', to construct generally covariant actions order-by-order in an expansion about an ordinary spacetime point. In section \ref{sec:dbs}, we show that our result (\ref{intact}) gives in a sense the result of carrying out their construction to all orders and provides an interpretation for the arbitrary coefficients encountered in their construction. 

The discussion in this paper makes no reference to any specific
scenario in string theory. We simply ask whether the group of
diffeomorphisms has a consistent representation on the combined space
of metrics and matrices\footnote{It is possible that the
diffeomorphisms are embedded nontrivially in some larger symmetry
group acting on this space. We comment briefly on this possibility in
section \ref{sec:additional}.} and whether we can find actions that
are invariant under the resulting transformations. Thus, our results
for D-particles should apply to D0-branes in any compactified type IIA
string theory or bosonic string theory, but also to the effective
actions describing pointlike branes arising from higher-dimensional
Dp-branes completely wrapped on p-cycles. Since the correct action may
be different in each of these situations, it is comforting that we
find a wide range of possible actions consistent with general
covariance. On the other hand, general covariance is not the only
principle that we can use to constrain the form of the action, and we
discuss some of the others in section 5. Finally, we offer a few
concluding remarks in section 6 and some technical results in the appendices.

The central ideas of this paper, representing the matrix degrees of freedom by a covariant matrix field on the transverse space and writing the action as an integral over the transverse space with a covariant distribution function, were motivated largely by the study of an analogous problem in \cite{blending}. There, the system under consideration was a set of pointlike branes inside a collection of higher-dimensional branes. In that case, the analogue of our diffeomorphisms is a local gauge symmetry associated with the gauge field on the higher-dimensional branes. The D-particle matrices themselves are invariant under this symmetry, but there are bifundamental fields, ``living'' at the matrix location of the D-particles, that transform non-trivially. In \cite{blending}, it was found that these bifundamental fields can be promoted to gauge covariant fields living on the higher-dimensional branes and that  gauge invariant actions for the bifundamentals may be written as integrals over the higher-dimensional branes with a matrix distribution function.

\section{Transformation rule for matrix coordinates}

We would like to write effective actions describing the coupling of multiple D-branes to a background metric such that the resulting actions are valid in any system of coordinates. For most of this work, we focus on the case of D0-branes described by matrix coordinates $X^i(t)$ on a curved space with metric $g_{ij}(y)$, since many of the essential difficulties are present already in this simple case. For now, we assume that the time-time and time-space components of the metric are trivial and consider only diffeomorphisms affecting the spatial metric.

\subsubsection*{The basic difficulty}

We begin by pointing out a basic difficulty with constructing generally covariant actions for multiple D-branes (many of these considerations were discussed previously in \cite{dbs}). Coordinate transformations are described by maps
\[
\tilde{y}^i = F^i(y)
\]
from one set of coordinates to another. Under such transformations, the metric transforms covariantly, 
\[
\tilde{g}_{ij} (F(y)) = \partial_{i} F^k \partial_{j} F^l g_{kl}(y) \;.
\]
The embedding coordinates $x^i(\sigma)$ of a single brane also have a simple transformation law
\be
\label{abeltrans}
\tilde{x}^i(\sigma) = F^i(x(\sigma)) \; .
\ee
In both cases, we have a local object (the metric at a point or the location of a point on the brane) whose transformation is determined by the value of the transformation function $F$ (or its derivative) at the point we are interested in. 

On the other hand, multiple D-branes, described by matrix coordinates
$X^i$, do not have well defined positions when the matrices don't
commute, but correspond to ``fuzzy'' higher-dimensional objects. As a
result, it is not at all obvious how the matrix coordinates should transform, since they are not associated with any single point or collection of points on the space. At a more technical level, the problem is that the abelian transformation law (\ref{abeltrans}) doesn't easily generalize to multiple D-branes, since we do not know what $F^i(X)$ means when $X$ is a matrix.

\subsubsection*{A naive approach and a consistency condition}

As a first attempt, we might try to generalize the transformation using a Taylor series,
\be
\label{naive}
X^i \to F^i[X] \equiv \sum_n {1 \over n!} \partial_{i_1} \cdots \partial_{i_n} F^i(0) X^{i_1} \cdots X^{i_n} \; .
\ee
For diagonal matrices, corresponding to branes with well defined positions, this gives the natural result that the coordinates of the individual branes transform via the abelian transformation law (\ref{abeltrans}). Further, this expression has no ordering ambiguities since the derivatives are automatically symmetrized. However, this result cannot be correct in general since it does not respect the group composition law for coordinate transformations. With the definition (\ref{naive}), it is straightforward to check that 
\be
\label{mismatch}
F[H[X]] \ne F \circ H [X]
\ee
where on the left side, we apply the transformation law (\ref{naive}) first with $H$, then with $F$, while on the right side we apply the transformation law (\ref{naive}) once using the composition of the two functions.

\subsubsection*{Transformation rule must involve the connection}

The first mismatch in (\ref{mismatch}) occurs at order $X^3$,
\be
\label{mism}
F[H[X]] - F \circ H [X] = {1 \over 12} \partial_i \partial_j F(H(0)) \partial_k \partial_l H^i(0) \partial_m H^j (0) [X^k,[X^l,X^m]] + {\cal O}(X^4) \; . 
\ee
To fix this, we can try to add additional terms at this order to the transformation 
law (\ref{naive}), with the requirement that all additional terms involve matrix commutators so as not to change the transformation law for diagonal matrices. In \cite{dbs},
it was shown that this procedure fails unless we allow the transformation law to depend explicitly on the metric; that is we must allow a transformation rule of the form
\[
X^i \to \tilde{X}^i = \Phi^i(F,X,g) \; .
\]
Allowing for this more general possibility, we find that it is possible to correct (\ref{naive}) order-by-order to achieve a transformation rule that respects the group composition law, which now takes the form\footnote{One might have required only that the two sides of this expression should agree up to a gauge transformation $\Phi \to U \Phi U^{-1}$. However, since we do not encounter any obstruction assuming that $U=1$, we will demand this more stringent requirement.}
\be
\label{comp}
\Phi(F, \Phi(H,X,g), \tilde{g}_H) = \Phi(F \circ H, X, g) \; .
\ee
For example, up to order $X^3$, we can take 
\bea
\Phi^i(F,X,g) &=& F^i(0) + \partial_j F^i(0) X^j + {1 \over 2} \partial_j \partial_k F^i(0) X^j X^k + {1 \over 3!} \partial_j \partial_k \partial_l F^i(0) X^j X^k X^l \cr
&& \qquad \qquad \qquad \qquad - {1 \over 12} \partial_m \partial_k
F^i(0) \Gamma^m_{jl}(0) [X^j,[X^k,X^l]]+ {\cal O}(X^4) \; .
\label{fixed} 
\eea
Here, $\Gamma$ is the Christoffel symbol associated with the metric $g$ (our conventions are summarized in appendix A). With this additional term, we find that the previous mismatch in (\ref{comp}) is resolved by the inhomogeneous term in the transformation law for the connection (given by (\ref{eqn:chr_xfm})).\footnote{In this paper, we consider only the metric connection, however, the mismatch will also be resolved by replacing $\Gamma$ by any more general connection with the same transformation rule.} 

In hindsight, since the matrices $X^i$ describe degrees of freedom distributed over some extended region of the curved space when they don't commute, it seems natural that 
the transformation law should involve the connection in this case.

\subsubsection*{Additional constraints}

An unsettling feature of (\ref{fixed}) is that the transformation law for $X$ seems to place special emphasis on the point $0$, the origin of our coordinates. On the other hand, in writing the rule (\ref{naive}) for diagonal matrices, we could have chosen any other point $y$ as the basis of our Taylor series, expanding in $X-y$ instead of $X$. Working order-by-order in $X-y$, we would then have found correction terms involving derivatives of the metric at $y$, needed to ensure the correct composition law at a given order, arriving eventually at a result
\be
\label{ytrans}
X^i \to \Phi^i(F(y), X-y, g(y))
\ee
analogous to (\ref{fixed}). Since there are no special values for the coordinates, the functional form of $\Phi$ for the correct transformation law should be independent of the choice of $y$. Furthermore, the expression on the right-hand side of (\ref{ytrans}) cannot depend on $y$, since the correct transformation rule cannot possibly depend on our choice of where to expand the functions. Thus, we must demand that  
\be
\label{yind}
{\partial \over \partial y^i} \Phi^j(F(y), X-y, g(y)) = 0 \; .
\ee
It is straightforward to check that this property is satisfied up to order $X^3$ using the $\Phi$ from our expression (\ref{fixed}). However, at the next order in $X$, there are transformation rules which satisfy the group composition law but which are not $y$-independent. Thus, the condition (\ref{yind}) must be imposed as an additional constraint on the transformation rule.

Finally, it seems natural to assume that for simple linear redefinitions of the coordinates (rotations, rescalings, shears, shifts), the matrix coordinates should transform in the same way, that is
\be
\label{linmap}
\Phi^i(M^k_l y^l+a^k, X, g) = M^i_j X^j + a^i \; .
\ee
This will be satisfied if all terms beyond those in the abelian expression (\ref{naive}) involve only second or higher derivatives on the transformation function $F$.

\subsubsection*{Summary of conditions on the matrix transformation rule}  

To summarize, we should demand that the transformation law for the matrix coordinates $X^i$ should be given by some expression
\be
\label{ytrans2}
X^i \to \Phi^i(F(y), X-y, g(y))
\ee
satisfying 
\begin{itemize}
\item Agreement with the abelian result (\ref{naive}) for diagonal matrices.
\item The composition law (\ref{comp}).
\item The coordinate independence constraint (\ref{yind}).
\item Natural behavior under linear redefinitions of coordinates (\ref{linmap}).
\end{itemize}

In addition to these, it appears consistent to assume one additional property for the transformation rule which holds trivially in the abelian case: 
\begin{itemize}
\item The transformation should be linear in the function $F$ and its derivatives:
\be
\label{linear}
\Phi^i(F(y)+ H(y), X-y, g(y)) = \Phi^i(F(y), X-y, g(y)) + \Phi^i(H(y),
X-y, g(y)) \; .
\ee
\end{itemize}
While this condition seems rather natural, we do not see a compelling reason to demand it, except that it will fix some ambiguity that would otherwise be present. On the other hand, we would be willing to abandon it (but not any of the others) if it turned out to be inconsistent with the remaining constraints. 

It is straightforward to verify that our order $X^3$ expression
(\ref{fixed}) satisfies all of these properties. Further, we have
checked that a solution exists to order $X^4$, with the result given
in appendix \ref{app:constraints}. In principle, we could continue order-by-order to
construct a consistent transformation law for $X$. Given this
transformation law, which is already rather unwieldy and
unenlightening at order $X^4$, we could try to construct an invariant
action (again order-by-order). Fortunately, there is a more elegant
approach, which we now describe. 

\section{A covariant object}   

Following \cite{blending}, we will proceed using the method familiar from gauge theory and general relativity: rather than working directly with $X$, we will look for a covariant object built from $X$, and then construct invariant actions using this. For the related problem considered in \cite{blending} this covariant object took the form of a field defined on the entire space transverse to the D-branes, such that the field evaluated at any point contained the same information as the original variables. It turns out that a similar construction is possible here.

\subsubsection*{The Abelian case: describing a point by a vector field}

To understand what sort of covariant object we should be looking for,
it is useful to step back to the case of a single brane. In this case,
the brane's configuration is simply a point $x^i$ on the transverse
space. Thus, the covariant object should be a field whose value at any
point $y$ encodes the information that the brane is at $x$. In fact,
there is a vector field that plays precisely this role, at least in
some neighborhood of the point $x$. The value of this field at any
point $y$ is a vector $v^i_y$ which is tangent to the geodesic between
$y$ and $x$, and has magnitude equal to the geodesic distance between
$y$ and $x$. In other words, to any point $x$ on the space, we
associate a section $v_y$ of the tangent bundle such that 
\be
\label{exp}
x = exp(v_y)
\ee           
where $exp$ is the exponential map. More explicitly, we have
\be
\label{vseries}
x^i = y^i + v^i_y - \sum_{n=2}^\infty \Gamma^i_{j_1 \cdots j_n} v^{j_1}_y \cdots v^{j_n}_y
\ee
where the $\Gamma$s are defined in appendix \ref{app:conventions}. The vector field $v_y$ is well defined globally for all points $x$ if and only if there exists a unique geodesic between any two points on the space. For now, we will be content to restrict to spaces of this type (or to demand that the D-branes are confined to regions with this property). In fact, as we discuss in section \ref{sec:discussion}, it is natural that our description should break down in more general cases.

Since our definition of $v_y$ is completely geometrical, it follows immediately that $v_y$ transforms covariantly as a vector field under diffeomorphisms 
\be
\label{vtrans}
\tilde{v}^{j}_{F(y)} = \partial_i F^{j}(y) v^i_y \; .
\ee 
Equivalently, if we introduce a vielbein $e_i^a(y)$, we can define the field $v^a_y = e^a_i(y) v^i_y$ with a tangent space index, so that $v^a_y$ transforms as a scalar under diffeomorphisms and as a vector under local Lorentz transformations. 

We now have two ways of describing the location of our single D-brane: either by specifying a point $x$ on the space, or by specifying a vector field whose exponential map is constant. While the latter may seem to be a rather perverse way of describing a single point, we note that the transformation law (\ref{vtrans}) generalizes much more easily in the case of matrix $V$ than the transformation law (\ref{abeltrans}) does for matrix $X$. In fact, (\ref{vtrans}) requires no modifications at all when $V$ is a matrix, since the transformation function $F$ is evaluated only at ordinary points $y$. 

\subsubsection*{A nonabelian generalization}

Thus, we are motivated to ask whether there exists some matrix valued field $V^i_y$ constructed from $X$ and the metric, such that under a change of coordinates, $V$ transforms covariantly as
\be
\label{vtrans2}
\tilde{V}^{j}_{F(y)} = \partial_i F^{j}(y) V^i_y  \; .
\ee
For diagonal $X$, $V^i_y$ should reduce to a diagonal matrix whose entries are vectors pointing along the geodesics connecting $y$ to the $N$ points $x^i$ describing the well-defined locations of the individual branes. Thus, the relation between $V$ and $X$ should agree with the abelian formula (\ref{vseries}) up to commutator terms. 

To determine if such a $V$ exists, the most straightforward approach would be to work order-by-order in $X$, via the following steps:
\begin{itemize}
\item Construct a consistent transformation rule for $X$ satisfying the constraints of section 2.
\item Write the most general expression for $V$ in terms of $X$ and the metric that agrees with the abelian result obtained by inverting (\ref{vseries}).
\item Determine whether the coefficients in the resulting expression may be chosen so that $V$ transforms as (\ref{vtrans2}) to the desired order.
\end{itemize}
Following these steps for the first several orders in $X$, one finds that it is possible to construct a covariant expression $V$ with the desired properties. This procedure quickly becomes tedious, however, but it turns out that there are a couple of better approaches. 

\subsubsection*{Direct construction from the transformation law}

First, it is not hard to show that given a consistent transformation rule for $X^i$, the existence of a covariant object $V_y$ follows immediately. For suppose that $X^i$ transforms via (\ref{ytrans2}) for some $\Phi$ satisfying all of our constraints. Intuitively, $\Phi$ gives us a rule to consistently promote any ordinary function $F^i(x)$ to a function on matrices. To define $V_y$, it turns out that we can simply apply this rule in the case where $F(x) = v_y(x)$, where $v_y$ is the ordinary function defined by inverting (\ref{vseries}). Thus, we take
\be
\label{vcanon}
V^i_y(X,g) \equiv \Phi^i(v_y(x,g),X,g).
\ee
To check that $V^i_y$ transforms covariantly, we note that
\begin{eqnarray}
V^j_{H(y)}(\Phi(H,X,g), \tilde{g}_H) &=& \Phi^j(v_{H(y)}(x,\tilde{g}_H),\Phi(H,X,g),\tilde{g}_H) \cr
&=& \Phi^j(v_{H(y)}(H(x),\tilde{g}_H) ,X, g) \cr
&=& \Phi^j(\partial_i H v_y^i(x, g),X,g) \cr
&=& \partial_i H^j(y) \Phi^i(v_y^i(x,g),X,g) \cr
&=& \partial_i H^j(y) V^i_y(X,g) \; .
\end{eqnarray}
Here, in going to the second line, we have used the composition law, in going to the third line, we have used the fact that $v$ transforms like a vector, 
\[
v_{H(y)}^i(H(x),\tilde{g}_{H}) = \partial_j H^i(y) v^j_y(x, g)
\] 
and in going to the fourth line, we have used the property (\ref{linmap}). Note that we do not require the linearity constraint (\ref{linear}). Thus, as long as a consistent transformation rule for $X$ exists, there exists a $V$ that transforms covariantly under it. As we will see below, this is certainly not the unique object with these transformation properties, but it represents a canonical choice.

\subsubsection*{Constructing the transformation rule from the covariant object}

The two approaches so far involve first constructing a consistent transformation rule for $X$ and then using this to find a $V_y$ which transforms covariantly. However, given the final relationship between $X$ and $V_y$,
\be
\label{any}
X^i  = y^i + \sum \Delta^i_{j_1 \cdots j_n}(g(y)) V^{j_1}_y \cdots V^{j_n}_y  
\ee
we could immediately go back and deduce the transformation rule for $X$, since we know how all quantities on the right hand side transform. In fact, since the transformation rules for $y$, $V_y$ and the metric $g$ all satisfy the composition law, it is automatic that the transformation rule for $X$ will, regardless of the form of the relationship (\ref{any}). The property (\ref{linmap}) is also guaranteed.

However, it is not true that any such relationship yields a well-defined transformation rule for $X$. The obstacle is that after transforming all quantities on the right hand side and eliminating $V_y$ in favor of $X$, the final expression for $X$'s transformation law may still depend on the arbitrary point $y$, and therefore fail to satisfy the property (\ref{yind}). By demanding that the final transformation rule is well defined (independent of the arbitrary point $y$), we are therefore able to constrain the relationship between $X$ and $V_y$. 

A further constraint comes if we demand the linearity relation (\ref{linear}). To ensure this, we must require that
\[
\tilde{\Delta}^i_{j_1 \cdots j_n} \partial_{k_1} F^{j_1} \cdots \partial_{k_n} F^{j_n}
\] 
is linear in $F$. This will be true as long as 
\be
\label{coeffconstr}
\Delta^i_{j_1 \cdots j_n} = T^i_{j_1 \cdots j_n} + \sum \Gamma^i_{j_1 \cdots j_m k_{m+1} \cdots k_{l}} (T_{\{nlm\}})^{ k_{m+1} \cdots k_{l}}_{j_{m+1} \cdots j_{n}}
\ee
where the $T$s are any tensor expressions built from the metric and the $\Gamma$s are the generalized Christoffel symbols defined in appendix A.

We thus have a third method for constructing a covariant $V_y$ which avoids first having to determine a consistent transformation rule:  
\begin{itemize}
\item Write the most general relationship of the form (\ref{any}) to some order in $V_y$ which reduces correctly to the abelian result for diagonal $X$ and which has coefficients of the form (\ref{coeffconstr}).
\item Transform all quantities on the right hand side.
\item Rewrite all occurrences of $V_y$ in terms of $X$ and the metric using the inverse of the relationship (\ref{any}).
\item Demand that the resulting transformation rule is independent of $y$ to determine the allowed coefficients $\Delta$. 
\end{itemize}

\subsubsection*{A differential equation for $V$}
 
In fact, we can go one step further and construct $V_y$ in terms of $X$ without ever having to refer to the transformation law. To see this, it is useful to go back and understand in another way why only certain choices for (\ref{any}) lead to a consistent transformation rule for $X$.

The point is that in general, it is not consistent to assume that $V_y$ is related to $X$ as in (\ref{any}) for all $y$ {\it and} that $V_y$ transforms as in (\ref{vtrans2}) for all $y$. If we assume that (\ref{any}) defines $V_y$ for all $y$, then by assuming that $V_y$ transforms as in (\ref{vtrans2}) for some particular $y$, we fix the transformation rule for $X$ as well as that for $V_{y'}$ for all other values $y' \ne y$. For a generic choice of (\ref{any}), this transformation rule for $V_{y'}$ will not agree with (\ref{vtrans2}) for $y' \ne y$. The ``good'' choices for (\ref{any}) are precisely those which imply that $V_y$ transforms as a vector field assuming that $V_y$ transforms as a vector some particular $y$. 

To translate this requirement into an explicit condition on (\ref{any}), it is enough to demand that $V_{y'}$ transforms as a vector for points $y'$ infinitesimally separated from $y$. To ensure this, we must demand that given the expression for $V_y$ in terms of $X$, that the covariant derivative of $V_y$ should be a tensor. Thus, we arrive at our final prescription to determine the allowed expressions for $V_y$ in terms of $X$ (and therefore the allowed transformation laws):
\begin{itemize}
\item Write the most general relationship of the form (\ref{any}) to some order in $V_y$ which reduces correctly to the abelian result for diagonal $X$ and which has coefficients of the form (\ref{coeffconstr}).
\item Invert the series to obtain an expression of the form
\be
\label{vany}
V^i_y = \sum V^i_{j_1 \cdots j_n} (g(y)) (X-y)^{j_1} \cdots
(X-y)^{j_n} \; .
\ee 
\item Take the covariant derivative of the right hand side with respect to $y$, and demand that the resulting expression is a tensor. That is, if we express the covariant derivative completely in terms of $V_y$ and expressions involving the metric (by eliminating $X$), only tensorial quantities should appear:  
\be
\label{covtens}
\nabla_j V^i_y = {\partial V^i_y \over \partial y^j} +
\Gamma^i_{jk}(y) V^k_y = T_j^i(V_y, g, R, \nabla R, \dots) \; .
\ee
\end{itemize}

Given any $V_y(X,g)$ that satisfies these constraints, we can if we wish evaluate explicitly the transformation rule for $X$, assured that the resulting rule will satisfy all of the requirements of section 2. 

\subsection{Solving the constraints}\label{sec:solving}

The constraints of the previous section can be explicitly implemented
to derive a set of recursion relations for the coefficients
$\Delta^i_{j_1 \cdots j_n}$ appearing in (\ref{any}). These are
derived explicitly in appendix \ref{app:constraints}. As we argue there, we may restrict to terms in (\ref{any}) which do not have $V$s contracted directly with the metric (more general solutions are possible, but we will not need them). 

With this restriction, we find that the most general solution (consistent with linearity of the transformation rule) to third order in $V$ is given by
\be
\label{thirdorder}
\Delta^i = V_y^i - {1 \over 2} \Gamma^i_{jk} V^j_y V^k_y - {1 \over
3!}( \Gamma^i_{jkl} + a R^i_{jkl} + b R^i_{kjl})V^j_y V^k_y V^l_y +
{\cal O}(V^4) 
\ee
where $a$ and $b$ are arbitrary coefficients. With this definition, the covariant derivative of $V$ is a tensor given by
\[
\nabla_i V_y^j = -\delta_i^j - [{1 \over 6}(1 - a - 2b) R^j_{kli}  +
{1 \over 6}(1 + a + 2b) R^j_{lki}] V_y^k V_y^l + {\cal O}(V^3) \; .
\]
Using this to evaluate the transformation rule for $X$, we find precisely the result (\ref{fixed}) with all dependence on the arbitrary coefficients dropping out.

\subsubsection*{Interpretation of arbitrary coefficients}

The arbitrary coefficients appearing in (\ref{thirdorder}) have a simple explanation.

Given some consistent transformation law for $X$, we should not expect there to be a unique object $V_y$ that transforms covariantly. Indeed, given any expression $V_y$ which transforms as a vector field, we may construct many other vector fields which reduce to the same quantity for diagonal $X$. We simply add quantities at higher orders built from curvature invariants and $V$s which vanish when the $V$s commute. For example, to third order in $V$, the most general possibility (again disallowing terms with $V$s contracted directly by the metric) is 
\be
\label{vredef}
(V^i_y)' = V^i_y + (\alpha R^i_{jkl}(y) + \beta R^i_{kjl}(y))V_y^j V_y^k V_y^l + \dots  \; .
\ee  
If $V_y$ transforms covariantly under some consistent transformation rule for $X$, then clearly $V'_y$ will transform covariantly under the same transformation rule. Thus, the arbitrary coefficients in (\ref{vredef}) will correspond to arbitrary coefficients in the general solution (\ref{thirdorder}) which do not affect the transformation law. Indeed, to third order in $X$, we have seen that there are precisely two such coefficients, and we now see that these correspond to the two arbitrary coefficients in (\ref{vredef}) at this order. 
 
It is an interesting open question whether all arbitrary coefficients
in the general solution for $V$ are related to redefinitions of the
form (\ref{vredef}). Since no coefficients of this type will affect
the transformation law, this assertion is equivalent to the statement
that there is a unique transformation law for the matrices $X^i$
satisfying all of our constraints. To test this, we have determined
the most general transformation law consistent with our constraints up
to order $X^4$. The result, given as equation (\ref{fourth}) in appendix \ref{app:constraints}, has 6 arbitrary coefficients. We emphasize that unlike the arbitrary coefficients in (\ref{thirdorder}) these appear directly in the transformation rule. There are a number of possible explanations for these:
\begin{itemize}
\item They are removed by constraints at higher orders. In other words, there may be only one choice for the coefficients that permits adding higher order terms to arrive at a consistent transformation law.
\item The various choices are related by some field redefinition of $X$ which preserves all of our conditions on the transformation law.
\item Our constraints are not stringent enough to uniquely specify the transformation law. In this case, either there could be additional constraints we have missed or there could be different inequivalent ways to represent the diffeomorphisms (perhaps appropriate for different types of D-particles). 
\end{itemize}
It will be interesting to see which of these possibilities is true, but we leave this as a question for future work.

\subsubsection*{A canonical choice for $V_y$}

We have seen that there will be infinitely many definitions of $V$ (corresponding to different physical quantities) covariant under the same transformation rule for $X$. However, there is one rather canonical choice, namely the quantity associated directly with the transformation law by the relation (\ref{vcanon}). At order $X^3$, this corresponds to the choice $a=b=0$ in the formula (\ref{thirdorder}) for which all  pure tensor terms beyond ${\cal O}(V)$ cancel. 

In fact, it is straightforward to show that the absence of pure tensor terms holds to all orders for this choice. To see this, we choose Riemann normal coordinates about the point $y$, with the result that (\ref{vseries}) becomes
\be
v_y(x) = x - y \; .
\ee
Inserting this into the definition (\ref{vcanon}), we find
\[
(V_y)^{RNC} = \Phi(F(x) = x-y,X,g) = X - y \; .
\]
Thus, our canonical choice for $V_y$ reduces simply to $X-y$ in Riemann normal coordinates, just as in the abelian case. If there were any higher order pure tensor terms, they would appear in any coordinate system, so we conclude that all of these (i.e. the coefficients $T$ in (\ref{coeffconstr})) must vanish for this choice.

For reference, we have listed the canonical expression for the relation between $\Delta$ and $V_y$ up to order $V^4$ in appendix \ref{app:constraints}. 

\section{Generally covariant actions}\label{sec:actions}

Having recast the information about our D-brane configuration into a covariant object, it will now be straightforward to construct invariant actions. 

We first note that any object constructed from a covariant $V_y$ together with tensors built from the metric, 
\[
{\cal L}(y) = {\cal L} (V_y, g(y), R(y), \nabla R(y), \cdots) 
\]
will transform as a scalar field under diffeomorphisms 
\[
\tilde{{\cal L}}(F(y)) = {\cal L}(y)
\]
as long as all indices are contracted. 

To obtain an invariant action from such a scalar, the standard procedure would be simply to integrate the scalar Lagrangian density over the space using the invariant measure $\sqrt{g} d^d y$. This sounds rather strange in our case, since the objects we are trying to describe are not spread over the entire space, but are typically localized in some region. Indeed, for the case of diagonal $X^i$, the branes sit at a collection of points, so we expect that the action should be a sum of terms corresponding to these points rather than an integral over the whole space. 

Actually, these considerations are consistent with an integrated action as long as our scalar Lagrangian density includes some kind of distribution function which localizes the integral to the vicinity of the branes. For diagonal $X^i$, this distribution function should reduce to a collection of delta functions, while in more general cases, we expect a smooth function which falls off away from the ``fuzzy'' brane configuration. 

\subsubsection*{Matrix distribution functions}

In fact, an object of this sort already appears in the actions describing the linearized couplings of branes to bulk fields in type II string theory expanded about flat space \cite{tv0,tvp,myers,oo}. For D0-branes, these couplings may be written
\bea
S &=& \int dt \sum_{n} {1 \over n!} \partial_{i_1} \cdots \partial_{i_n} b_{\alpha}(0) {\rm Str}( {\cal O}^\alpha(X, \dot{X}) X^{i_1} \cdots X^{i_n}) \cr
&=& \int dt \; \int d^9 y \; b_{\alpha} (y) \; {\rm Str}( {\cal O}^\alpha(X, \dot{X}) \delta(X-y))
\eea
where $b$ is some bulk field, ${\rm Str}$ denotes the symmetrized trace,\footnote{When ${\cal O}$ is a product of terms, the symmetrized trace averages over all orderings of the individual terms with the other $X$s coming from the delta function. Commutators appearing in ${\cal O}$ are symmetrized as a unit.} ${\cal O}$ is an operator constructed from $X$ and $\dot{X}$, and $\delta$ is a matrix distribution function defined as
\be
\label{delta}
\delta(X-y) = \int {d^9 k \over (2 \pi)^9} \; e^{i k \cdot (X-y)} \; .
\ee
For diagonal $X$, this reduces to a diagonal matrix of $\delta$ functions, so that the Lagrangian reduces to a sum of terms corresponding to the individual branes. For more general configurations, the support of $\delta(X-y)$ is no longer a collection of points, but some higher-dimensional extended region, as desired (for a recent discussion of this object, see \cite{hashimoto}).

\subsubsection*{A covariant distribution function}

As it is, the matrix distribution function (\ref{delta}) is not appropriate for inclusion in our covariant action, since it has a very complicated transformation law inherited from the transformation rule for $X$. On the other hand, if we promote $X-y$ to its covariant generalization $V_y$, we immediately obtain a distribution function that transforms as a scalar field, as long as we choose the index on $V_y$ to be a tangent space index rather than a spacetime index. That is, we define
\be
\label{covdist}
\delta(V_y) = \int {d^d k \over (2 \pi)^d} \; e^{i k_a V^a_y} = \int
{d^d k \over (2 \pi)^d} \; e^{i k_a e^a_i(y) V^i_y} \; .
\ee
This transforms as a scalar under diffeomorphisms since $V^a_y$ is a scalar, while under local Lorentz transformations the expression is also invariant since
\[
\delta(V_y) \to \int {d^d k \over (2 \pi)^d} \; e^{i k_a \Lambda^a {}_b(y) V^b_y} = \int {d^d \tilde{k} \over (2 \pi)^d} \; e^{i \tilde{k}_b V^b_y} = \delta(V_y)
\]
where we have used $|\det(\Lambda)|=1$.  If desired, we can rewrite the matrix distribution function (\ref{covdist}) directly in terms of the vector field $V^i_y$ by absorbing the vielbein into the integration variable $k$. This gives
\[
\delta(V_y) = {1 \over \sqrt{g(y)}} \int{d^d k \over (2 \pi)^d} \; e^{i k_i V^i_y}
\]
where $i$ is a spacetime index. 

\subsubsection*{Result for generally covariant actions}

Using this covariant distribution function, we may now write our final result for the form of possible generally covariant actions describing multiple D0-branes coupled to a metric,
\be
\label{result}
S = \int dt \int d^d y \sqrt{g(y)} \; {\rm Tr } \left( {\cal L}(V_y, g, R, \nabla R, \dots) \delta(V_y) \right) \; .
\ee
We emphasize that there are an infinite number of possibilities here, since ${\cal L}$ can be an arbitrary function of $V$ and the curvatures. In fact, we will demonstrate shortly that {\it any} generally covariant action may be written in this form.\footnote{This justifies making one specific choice for the matrix distribution function even though there may be other possible definitions with the same transformation properties. Also, there is no need to consider more general ordering prescriptions such as the symmetrized trace, although the choice (\ref{result}) may not yield the simplest expression for a particular action.} 

\subsubsection*{Expansion about a point}

The integrated form of the action above is completely natural given that general D-brane configurations correspond to fuzzy higher-dimensional objects which should couple to the metric over some region of space rather than at a finite collection of points. However, when expanding about a configuration in which all branes are coincident at a single point (which we may take to be $y=0$), it is useful to have an explicit form for the action as a series in powers of $X$. To obtain this, we define (all derivatives are with respect to $y$)
\be
\label{defl}
{\cal L}_{i_1 \cdots i_n} = {1 \over n!} \partial_{i_1} \cdots \partial_{i_n} {\cal L}|_{y=0}
\ee
and 
\be
\label{defw}
W^i_{i_1 \cdots i_n} = {1 \over n!} \partial_{i_1} \cdots \partial_{i_n} (V^i_y + y^i)|_{y=0}
\ee
so that 
\[
S = \int dt \int d^d y \; {\rm Tr} \left( \left\{{\cal L}|_{y=0} + {\cal L}_i y^i + {\cal L}_{ij}
y^i y^j + \dots \right\} e^{-(W^i + W^i_j y^j + W^i_{j k} y^{j} y^{k}
+ \dots) { \vec{\partial} \over \partial y^i}} \delta(y) \right) \; .
\] 
Here, the derivative in the exponential acts only on the delta function. To evaluate this, we expand the exponential, keeping only terms with an equal number of $y$s and derivatives. We find that
\be
\label{actexp}
S = \int \; dt \sum {\rm Tr} \left( {\cal L}_{i_1 \cdots i_n} \{ W^{(i_1} \cdots W^{i_n} W^{j_1} \cdots W^{j_m)} \}_{j_1 \cdots j_m} \right)
\ee
where the sum is over all distinct choices for how to distribute the indices $\{ j_1, \cdots, j_m \}$ (in order) over the various $W$s.\footnote{There are $(n+2m-1)!/(m!(n+m-1)!)$ terms for given $n$ and $m$.} Here, a term with ${\cal L}_{i_1 \cdots i_n}$ and $k_m$ occurrences of $W$ with $m$ lower indices will have terms starting at order $X^{c_n + n + k_1 + k_2 + 2k_3 + \dots}$ where $c_n$ is the power of $X$ appearing in the leading term in ${\cal L}_{i_1 \cdots i_n}$. In particular, at each order in $X$ there will be some finite number of terms. 

\subsubsection*{Minimal basis of invariant actions}

Given the apparently large number of covariant actions, it is useful to have some way to classify the various possibilities. Since the set of invariant actions forms a vector space, a convenient method of characterization will be to describe a minimal basis, such that the most general invariant action is some linear combination of the basis elements. 

First, we consider any generally covariant action, expanded in powers of $X$ around the configuration $X=0$.\footnote{For this discussion, we restrict to actions which admit an expansion in powers of the fields, though there are more general possibilities (see e.g. \cite{odb}, section 9.1).} Let the terms in the Lagrangian at leading order in $X$ be denoted $L_0$, and suppose that $n$ is the power of $X$ that appears. Now, the full action is invariant under coordinate transformations, so in particular, it must be invariant under an infinitesimal coordinate shift $y \to y + \epsilon$. Under such a transformation, the variation of $L$ contains terms of order $X^{n-1}$ and higher, with the $X^{n-1}$ terms coming exclusively from the variation of $X$s appearing in $L_0$. The condition that these terms vanish is therefore equivalent to the condition that 
\be
\label{cond1}
\partial_{X^i} L_0 \equiv \partial_{\epsilon^i} L_0(X + \epsilon) = 0
\; .
\ee
This will be satisfied, for example, if all $X$s appear in commutators or with time derivatives. 

Next, consider any finite coordinate transformation $y \to F(y)$ such that $F(0) = 0$. Under such a transformation, we have
\[
X^i \to \partial_j F^i(0) X^j + {\cal O}(X^2) \; .
\] 
Thus, the leading order terms in the transformed action will come exclusively from the leading order terms in the original action. Now, consider the collection of all leading terms with a given cyclic ordering of $X$s and $\dot{X}$s in the trace,
\[
A_{i_1 \cdots i_n} T^{i_1 \cdots i_n}
\]
where $A$ will be some function of the metric and its derivatives at the point $y=0$ and $T$ indicates the trace of some product of $X$s and $\dot{X}$s. Under the coordinate transformation above, we have
\[
\partial_{j_1} F^{i_1}(0) \cdots \partial_{j_n} F^{i_n}(0) \tilde{A}_{i_1 \cdots i_n} T^{j_1 \cdots j_n} + {\cal O}(X^{n+1}) \; .
\]
Since there will be no other terms in the transformed action proportional to $T^{j_1 \cdots j_n}$, it must be that 
\[
\partial_{j_1} F^{i_1}(0) \cdots \partial_{j_n} F^{i_n}(0) \tilde{A}_{i_1 \cdots i_n}  = A_{j_1 \cdots j_n} \; ,
\]
in other words, $A$ must be a tensor. 

Thus, we have shown that the leading order terms in any invariant action must involve the metric only in tensors and must involve $X$ only in derivatives and commutators (or more precisely, satisfy (\ref{cond1})). We will now show that any such term can be extended with the addition of higher order terms to give an invariant action. 

For suppose that $L_0 = \tr({\cal L}(X, \dot{X}, g(0), R(0), \nabla
R(0), \dots))$ satisfies (\ref{cond1}), where ${\cal L}$ is a scalar. Then we argue in appendix \ref{app:technical} that there must be some ${\cal L}'$ such that $\tr({\cal L}') = \tr({\cal
L})$ and
\be
\label{cond2}
\partial_{X^i} {\cal L}' = 0 \; .
\ee
That is, by some equivalent rearrangement of terms in the trace, we
can ensure that (\ref{cond1}) is satisfied without having to take the
trace.\footnote{This is not true for certain scalar terms at order $X^3$ involving background fields other than the metric. For an example, see the discussion of couplings to other fields below.}

Now, the Lagrangian 
\be
\label{covext}
L = \int d^d y \sqrt{g(y)} \; {\rm Tr } \left( {\cal L}'(V_y,\dot{V_y}, g(y), R(y), \nabla R(y), \dots) \delta(V_y) \right)
\ee
is covariant, by our construction. The expansion of this Lagrangian in powers of $X$ is given by our result (\ref{actexp}) from the previous section. If ${\cal L}'$ is order $n$ in $X$, the condition (\ref{cond2}) ensures that none of the expressions ${\cal L}_{i_1 \cdots i_n}$ (defined in (\ref{defl})) have any terms at lower order than $X^n$. As a result, the only term in (\ref{actexp}) which contributes at order $n$ is the first term 
\be
  \tr({\cal L}'(V_0,\dot{V_0}, g(0), R(0), \nabla R(0), \dots))
= \tr({\cal L}(V_0,\dot{V_0}, g(0), R(0), \nabla R(0), \dots)) \; .
\ee
Finally, since $V_0 = X + {\cal O}(X^2)$, the order $X^n$ term in the full action (\ref{covext}) is exactly the leading order expression $\tr({\cal L}(X, \dot{X}, g(0), R(0), \nabla R(0), \dots))$ that we started with. 

It now follows immediately that a minimal basis of generally covariant actions may be obtained by choosing a minimal basis of terms ${\cal L}_I$ built from tensors and satisfying (\ref{cond1}) and then taking the covariant extensions (\ref{covext}). For if $L$ is any invariant Lagrangian, we can certainly construct a Lagrangian $L'_1$ using our basis with the same leading terms, such that $L-L_1'$ is a higher order invariant Lagrangian. We can then construct a Lagrangian $L'_2$ using our basis that reproduces the leading terms in $L-L_1'$, and so forth, so that finally we obtain $L = \sum_k L'_k$.

\subsection{Relation to the base-point independence approach of de Boer and Schalm}\label{sec:dbs}

In our discussions above, we have built covariant actions by constructing a Lagrangian density ${\cal L}(y)$ which transforms as a scalar field and integrating this over space with an invariant measure (and a scalar distribution function). It is possible, however, that we could avoid having to integrate if it happens that ${\cal L}$ itself is independent of $y$. In this case, under a coordinate transformation we would have
\[
\tilde{\cal L}(y) = {\cal L} (F^{-1}(y)) = {\cal L}(y)
\]    
so ${\cal L}$ itself would be an invariant action. This observation suggests the following alternative procedure for constructing invariant actions:
\begin{itemize}
\item Start with some scalar expression ${\cal L}_0 (V_y,g(y), R(y), \cdots)$ at a given order in $V_y$.
\item Calculate $\partial_{y^i} {\cal L}_0$.
\item Try to add a higher order scalar expression ${\cal L}_1$ so that the leading terms in $\partial_{y^i} {\cal L}_1$ cancel those in $\partial_{y^i} {\cal L}_0$.
\item Continue adding terms at higher orders to eventually arrive at an expression $\sum_n {\cal L}_n$ that is $y$-independent and therefore generally covariant.
\end{itemize}  
It turns out that this procedure is precisely equivalent to the
``base-point independence'' method proposed by de Boer and Schalm
\cite{dbs}. Their discussion did not explicitly involve a covariant
field $V_y$, but rather the assumption that one was working in Riemann
normal coordinates about the point $y$, and therefore that the action
could be written directly in terms of $X$ and tensors built from the
metric. However, as we saw in section \ref{sec:solving}, it is possible to make a choice of $V_y$ that reduces simply to $X-y$ in Riemann normal coordinates about the point $y$. With this choice, the procedure we have just outlined is identical to that of de Boer and Schalm, except in coordinate-independent language.\footnote{The authors of \cite{dbs} also considered constraints beyond base-point independence to further constrain their actions. The analogues of these additional constraints in our language are discussed in section 5, but from our perspective, these appear to go beyond the requirements of diffeomorphism invariance.}

At first sight, it sounds rather unlikely that this method will work, since in general there is no guarantee that higher order terms exist to cancel the leading $y$-dependence of a given scalar operator. However, we now show that it will work precisely when the scalar action ${\cal L}_0$ we start with satisfies the condition (\ref{cond1}), and that the general result of carrying out the base-point independence procedure to all orders is exactly the action 
\be
\label{dbssol}
\int dt \int d^d y \; \sqrt{g} \; {\rm Tr} (({\cal L}_0 + {\cal L}')\delta(V_y))  
\ee
where ${\cal L}'$ is some arbitrary scalar Lagrangian density whose leading terms are at higher orders than those of ${\cal L}_0$.

To see this, we must show that the integrated actions are equivalent to some $y$-independent scalar function of $V_y$ and tensors built from the metric. Consider then any covariant action defined by an integral expression (\ref{result}) and expanded about $y=0$ in powers of $X$ as described above. If we now go to Riemann normal coordinates about the point $y=0$, then it will be possible to rewrite the expanded action entirely in terms of tensors built from the metric (as well as $X$ and $\dot{X}$). Further, as discussed in section \ref{sec:solving}, there is some choice of $V_y$ for which $V_0 = X$ in Riemann normal coordinates about the point $y=0$. Making this choice, we may replace $X$ with $V_0$ everywhere in our expanded action, to obtain an expression for the Lagrangian
\[
L = L(V_0, \dot{V}_0, g(0), R(0), \dots)  \; .
\]
This expression was derived in Riemann normal coordinates, but is now written completely in terms of tensors evaluated at the point $y=0$. Therefore, under a coordinate transformation that maps $0$ to some other point $y$ this maps to 
\[
L' = L(V_y, \dot{V}_y, g(y), R(y), \dots)  \; .
\]
But since we started with an invariant Lagrangian, we must have $L'=L$, so the scalar expression $L(V_y, \dot{V}_y, g(y), R(y), \dots)$ must be $y$-independent. 

Applying this procedure to (\ref{dbssol}), we therefore obtain a
$y$-independent scalar expression whose leading term (assuming that ${\cal L}_0$ satisfies (\ref{cond1})) is ${\cal L}_0 (V_y,g(y), R(y), \cdots)$ as desired. Thus, our integral expressions represent a closed form solution to the base-point independence constraints. Since (\ref{dbssol}) should be the most general solution to the base-point independence constraints whose leading terms match ${\cal L}_0$, the arbitrary coefficients found in \cite{dbs} should correspond to the arbitrary higher order scalar terms ${\cal L}'$ in (\ref{dbssol}). 

Since the method we have described apparently will give results in complete agreement with the base-point independence method of \cite{dbs}, it is useful at this point to highlight a few important differences and how we feel that the current approach goes beyond the previous work. 
\begin{itemize}
\item
In this work, the problem of determining generally covariant actions is reduced to determining an expression for our covariant matrix field $V_y$ (or equivalently, determining how the matrices $X^i$ transform under diffeomorphisms). While we currently do not have a closed form solution for $V_y$ and thus must resort to an order by order analysis similar to that in \cite{dbs}, this need only be done for the single object $V_y$ after which the most general coordinate-independent action may be written in closed form as (\ref{result}). Thus, we are able to treat at once all possible covariant actions (by describing a minimal basis) starting at any order in derivatives. 
\item
In this work, we have determined the general solution to the constraints of demanding invariance under ordinary  diffeomorphisms, without any other conditions. In particular, our results are not specific to D0-branes in non-compact type IIA string theory, but are equally valid for any system of pointlike D-branes. On the other hand, by requiring certain additional constraints (to be discussed below), the authors of \cite{dbs} arrive at a result for the type IIA D0-brane action with fewer undetermined coefficients than the action we obtain by diffeomorphism invariance alone.
\item
Our integral form (\ref{result}) of the action has manifest general covariance, since it is written completely in terms of tensorial quantities with simple transformation properties. Further, we feel that writing the action as an integral over the transverse space (rather than an expansion about some specific point) is the most natural way to describe the coupling of the metric to objects which are not necessarily localized near any single point. For example, using an action expanded about some particular point, the effects of any feature of the metric at some finite distance away from our chosen point will be incorporated correctly only by including an infinite number of terms in the action.
\end{itemize}

\subsection*{Examples}

To conclude this section, we describe the leading order terms in the most general actions involving 0 and 2 time derivatives. We restrict to terms of the minimum possible scaling dimension, and further assume that the background satisfies its equations of motion $R_{ij} = 0$ (appropriate if we want to use the action to study branes on some fixed background geometry). 

With these restrictions, the most general invariant potential term is 
\beas
L_{potl} &=& \int d^d y \; \sqrt{g(y)} \; \tr(\delta(V_y)\; \{
\frac{1}{4} g_{ij}(y) g_{kl}(y) [V_y^i , V_y^k ][V_y^j, V_y^l] \cr
&& +  g_{ij} R_{klmn} (a [V_y^i,V_y^k][V_y^j,V_y^n][V_y^l,V_y^m] + b [[V_y^k,V_y^l],V_y^i][[V_y^m,V_y^n],V_y^j])\}) \; .
\eeas
Note that in adding the arbitrary higher order terms, we need only consider independent terms satisfying (\ref{cond1}). Writing this explicitly as a $y$-independent tensor expansion, we find:
\bea
\label{genpotl}
L_{potl} &=& \tr ( \frac{1}{4} g_{ij} g_{kl} [V^i , V^k ][V^j,
V^l] \\
&+& g_{ij}R_{klmn} \left( 8b V^i V^j V^k V^n V^l V^m +(a-8b+ \frac{1}{4}) V^i V^j V^k V^n V^m V^l  \right. \cr
&&   +  (a+{1 \over 12})V^i V^k V^j V^n V^l V^m 
 - (2a + {1 \over 6}) V^i V^k V^j V^n V^m V^l \cr 
 && \left.  -(a+8b+{1 \over 12}) V^i V^k V^n V^j V^l
V^m +  (a+8b-{1 \over 12})V^i V^k V^n V^j V^m V^l \right) +
{\cal O}(V^7)) \; .
 \nonumber
\eea
Thus, general covariance determines the six coefficients at order $X^6$ up to two arbitrary parameters.

For the kinetic term, we have 
\[
L_{kin} = \int d^d y \; \sqrt{g(y)} \; \tr(\delta(V_y)\; \{
\frac{1}{2} g_{ij}(y) \dot{V}^i_y \dot{V}^j_y +  R_{ijkl} (c \dot{V}_y^i \dot{V}_y^l [V_y^j, V_y^k] + d
[\dot{V}_y^i,V_y^j] [\dot{V}_y^l,V_y^k]) \} ) \; .
\]
Writing this explicitly as a $y$-independent tensor expansion, we find:
\bea
\label{genkin}
L_{kin} &=& \tr( \frac{1}{2} g_{ij} \dot{V}^i \dot{V}^j + R_{ijkl} \{ ({1 \over 4} + c) \dot{V}^i \dot{V}^l V^j V^k - ({1 \over 4} +c+2d) \dot{V}^i \dot{V}^l V^k V^j  \cr 
&& \qquad \qquad + ({1 \over 12} + d) \dot{V}^i V^j \dot{V}^l  V^k + ({1 \over 12}+d) \dot{V}^i V^k \dot{V}^l V^j \} + {\cal O}(V^5)) \; .
\eea

If desired, these terms may be written out explicitly in terms of our original matrix $X$ using the relation (\ref{fourthorder}) between $X$ and $V$. We note in particular that the lowest order (in $X$) correction terms necessary to covariantize a given leading order term are uniquely determined, since they arise from the uniquely determined order $X^2$ term in the definition of $V$. 

\section{Additional constraints}\label{sec:additional}

We have seen that the constraints of ordinary general covariance leave a large variety of possible actions. This should be expected, since there are many different situations to which our analysis should apply, for example any string theory compactification which contains D-branes that are pointlike in the non-compact directions. On the other hand, there should be additional constraints that restrict the action further, both in general and in particular situations. We mention a few of these (discussed previously by \cite{dbs,dbs2,douglas2,douglas,tvp,myers} and others) presently.

\subsubsection*{Emergence of $V_y$ from the geodesic equation}

In the abelian case, the relation between $v_y$ and $x$ follows directly from the geodesic equation. That is, choosing initial conditions $x^i(0) = y^i$ and $\dot{x}^i(0) = v^i$, we obtain a solution $x^i(t,v,y)$ such that $x^i(1,v,y)$ reproduces the relation (\ref{vseries}) between $x$ and $v$. It is natural to suspect that the correct kinetic term in the nonabelian case shares this property. 

That is, if we write the matrix geodesic equation obtained by varying the kinetic term with respect to $X^i$, the solution with $X(0) = y$ and $\dot{X}(0) = V_y$ should have $X(1,y,V_y)$ equal to the relationship (\ref{fourthorder}) between $X$ and $V_y$ (for the canonical choice of $V_y$). An equivalent requirement is that if we vary the action with respect to $V_y$ for some choice of $y$ then $V_y^i = A^i t$ should be a solution with any constant matrix $A$. 

Equivalently, since $V^i_0 = X^i$ in Riemann normal coordinates for our canonical choice of $V$, we can demand that in Riemann normal coordinates, $X^i = X^i_0 t$ is a solution to our matrix geodesic equation. In this form, the condition is identical to an extra constraint imposed by de Boer and Schalm in their construction of a D0-brane kinetic term. 

It is easy to check that this condition will be satisfied for our
general kinetic term (\ref{genkin}) if we fix the arbitrary
coefficients to satisfy $c+4d=-5/12$. 

\subsubsection*{T-duality}

For D-particle actions arising in string theory, the potential and
kinetic terms should be related to each other by T-duality (see
\cite{tvp,myers} for extensive discussions of the constraints placed
on D-brane actions by T-duality). In the simple case where the time
direction is an isometry direction and all components of the metric
involving time are trivial, the kinetic terms in the Lorentzian D-particle action (up to an overall minus sign) follow from terms involving $X^0$ in the D-instanton action via the replacement
\be
\label{replace}
[X^0, X^i] \to D_0 X^i \; .
\ee
Further, the D-instanton action should be identical in form (again up to an overall minus sign) to the
potential terms in the D-particle action since these arise from the
subset of terms in the D-instanton action not involving $X^0$. Thus,
if the constraints of T-duality apply, the kinetic terms may be
formally obtained from the potential terms by allowing all indices in
the potential term to run from $0$ to $d$ (which gives the D-instanton
action), setting $g_{00}=1$ and all other tensors with a $0$ index to
zero, and making the replacement (\ref{replace}). For the leading
potential and kinetic terms (\ref{genpotl}) and (\ref{genkin}), the
constraints of T-duality imply that $a=c$ and $4b=-d$.

\subsubsection*{The geodesic distance criterion}

In \cite{douglas2,douglas}, Douglas suggested a number of conditions that should be satisfied by actions describing multiple D-branes on a curved space. While he did not explicitly include the requirement of general covariance, a related constraint was provided by his ``geodesic distance criterion''. This states that when expanding the action about a diagonal matrix configuration (corresponding to branes with well-defined positions), the physical fluctuations corresponding to off-diagonal matrix elements $y_{ab}$ (which arise from strings stretched between the branes) should have masses proportional to the geodesic distance $d_{ab}$ between the corresponding branes (at $x_a$ and $x_b$). 

Since the actions we have described are generally covariant, it
follows that the masses of off-diagonal modes must be given by some
coordinate independent quantity that reduces to the coordinate
distance in flat space (assuming we are using a covariant
generalization of the usual kinetic term and commutator-squared
potential). While the geodesic distance is one such quantity, it is
not the unique quantity with this property (for example, we may insert
curvature invariants into the integral defining the geodesic
distance). Thus, there may be sensible generally covariant D-brane actions that do not satisfy the geodesic distance criterion (for a possible example, see \cite{dos}).

To determine what constraints the geodesic distance condition places on our leading order actions, note that the mass matrix coming from the general potential and kinetic terms in (\ref{genpotl}) and (\ref{genkin}) is
\be
\label{massmat}
{\cal M}_{ij} = d^2_{ab} ( \delta_{ij} - \hat{d}^i_{ab} \hat{d}^j_{ab} + (2d +
8b) R_{iklj} d^k_{ab} d^l_{ab} + \dots)
\ee
where $\hat{d}^i_{ab} = d^i_{ab} / \sqrt{d_{ab}^2}$.  The geodesic
distance criterion thus holds (at this order) if and
only if we choose $4b=-d$, which is necessarily true if we satisfy the T-duality constraint.\footnote{Actually, the geodesic distance criterion may hold more generally if the space is assumed to be Ricci-flat (satisfying the bulk equations of motion). In this case, the order $R$ contribution to the eigenvalues of the mass matrix will vanish, since it necessarily involves the Ricci tensor (coming from the trace of the Riemann tensor appearing in (\ref{massmat})). The first non-vanishing contribution will be at order $R^2$, and this may cancel with contributions from explicit $R^2$ terms in the action that we have not yet considered.} 
 
In \cite{dbs}, the authors argued that the geodesic distance criterion was satisfied exactly for the class of actions they considered. In our language, these actions should be generally covariant, and satisfy the additional constraints of the previous two subsections: that $X(V)$ should arise from a solution of the matrix geodesic equation, and that the potential terms should take a specific form related to the kinetic term by T-duality. Thus, while the geodesic distance criterion provides a constraint beyond that of general covariance it seems to be implied by the extra constraints we have outlined already. 

\subsubsection*{Agreement with known results}

So far, we have not made reference to any specific scenario in string theory. However, given a specific example of D-particles (for example D0-branes in bosonic string theory, or fully-wrapped D4-branes in type IIA string theory on K3), we should of course demand that the action agrees with any known results.  

As an example, we can look at the most well-known occurence of D-particles, the D0-branes in noncompact type IIA-string theory. For this case, many terms in the effective action are known \cite{myers,tv0,tvp}, and the correct covariant action should agree with these. For example, the leading terms coupling D0-branes to a weak transverse metric $g_{ij} = \eta_{ij} + h_{ij}$ are \cite{tv0}
\[
S = \int dt \int d^9 y {1 \over 2} h_{ij}(y) {\rm Str} ((\dot{X}^i \dot{X}^j + [X^i,X^k][X^j,X^k])  \delta(X-y)) \; .
\] 
Demanding that these terms are reproduced by the actions
(\ref{genkin}) and (\ref{genpotl}) fixes the coefficients in those
expressions to be $c=-7/36, d=-1/18$ and $a=-7/36 ,b=1/72$ (though it does
not fix all arbitrary coefficients at higher orders \cite{dbs}).
These values are consistent with all of the constraints of the previous three subsections. 

In this case, general covariance constrains the form of the terms non-linear in the metric given the linearized results. Such constraints will apply also to terms involving other background fields. For example, the linearized coupling 
\[
\int dt \int d^9 y C_0(y) {\rm Tr} ( \delta(X-y))
\]
of D0-branes to the time component of the Ramond-Ramond one-form field must generalize to some covariant expression
\[
\int dt \int d^9 y \sqrt{g(y)} C_0(y) {\rm Tr} (\delta(V_y) (1 + {\cal L'}(V_y, g)))
\]
where ${\cal L'}$ is some higher order scalar commutator expression.  Another example is provided by a covariant generalization of the coupling responsible for the Myers effect \cite{myers}. Here, the linearized coupling is proportional to \cite{tv0}
\be
\label{myers}
\int dt \int d^9 y \partial_i C_{0jk} \tr(X^i[X^j,X^k])
\ee
which must generalize to 
\[
\int dt \int d^9 y \sqrt{g(y)} C_{0jk}(y) {\rm Tr} ( \delta(V_y) ([X^j,X^k] + {\cal L'}(V_y, g))) \; .
\]
In this case, it is the Ramond-Ramond potential that appears in the covariant action, rather than its derivative, and the term (\ref{myers}) arises from the ${\cal L}_i W^i$ term in (\ref{actexp}).

\subsubsection*{A larger group of symmetries?}

In this paper, we have considered only the ordinary group ${\cal D}$ of diffeomorphisms, asking whether this group can be represented on the space of metrics and D-particle matrices. It is possible, however, that there is some larger group ${\cal D}^+$ of symmetries acting on this space such that the diffeomorphism group arises upon restriction to diagonal matrices. In this case, for a given element $F^+ \in {\cal D}^+$
\[
F^+ : (X,g(y)) \to (\tilde{X}, \tilde{g}(y))
\] 
there should be an element $\pi(F^+) \in {\cal D}$ such that 
\[
F^+ : (diag(x_1, \dots x_n),g(y)) \to (diag(\pi(F^+)(x_1), \dots \pi(F^+)(x_n)), \tilde{g}_{\pi(F^+)}(y)) \; .
\]
It follows automatically that 
\[
\pi(F^+ \circ H^+) = \pi(F^+) \circ \pi(H^+)
\]
so $\pi$ would define a homomorphism between ${\cal D}^+$ and ${\cal D}$. 

In the general situation, there could be many elements in ${\cal D}^+$ which reduce to a given element of ${\cal D}$. Thus, it is not clear a priori that there should be any canonical way to associate an element of ${\cal D}^+$ to a given diffeomorphism as we have demanded in this paper. Requiring the existence of a consistent map
\be
\label{map}
X \to \Phi(F,X,g(y))
\ee
satisfying the group composition law, as we have done in here, amounts to requiring that there is some homomorphism acting in the opposite direction from ${\cal D}$ to ${\cal D}^+$. This map would certainly be one-to-one so it would be an isomorphism between ${\cal D}$ and some subgroup of ${\cal D}^+$. 

The fact that we do seem to find a consistent map (\ref{map}) (at least to order $X^4$) may therefore be interpreted as evidence that if some larger group of symmetries ${\cal D}^+$ exists, it should have at least one subgroup isomorphic to the diffeomorphism group. It should be interesting to investigate whether or not such a larger group of symmetries exists, and if so, to understand the additional constraints on the action arising from the requirement of invariance under this group.\footnote{Of course, we have in addition the worldvolume gauge symmetry $X(t) \to U(t) X(t) U^{-1}(t)$, so at the least we have a product of this symmetry group with the diffeomorphisms. It is also possible that this symmetry is mixed nontrivially with the diffeomorphisms in some larger group. Relatedly, in appendix \ref{app:promote}, we note that the worldvolume gauge symmetry itself can be extended to a local spacetime symmetry via the introduction of a flat connection in addition to our field $V_y$.}

\subsubsection*{Non-transverse metrics and worldvolume reparametrization invariance}

We have focused here exclusively on target space diffeomorphisms involving the transverse metric. Of course, there are additional diffeomorphisms that mix the transverse coordinates with those in the brane directions (e.g. the time direction for D-particles). A fully geometrical action for multiple D-branes should be invariant under these as well, and implementing this symmetry should lead to additional constraints on the action. This issue was discussed in detail in \cite{dbs2}. As noted by these authors, implementing the more general diffeomorphisms is hindered by the fact that  the transverse D-brane coordinates, described by matrices, are treated very differently from the others, which have been identified with the worldvolume coordinates in the usual description. 

To implement full diffeomorphism invariance, the authors of \cite{dbs2} suggest that there should be a more democratic description, with matrix coordinates $X^\mu$ for all directions, plus some enhanced version of worldvolume reparametrization symmetry that permits choosing a nonabelian version of the static gauge $X^a(\sigma) = \sigma^a \identity$. Combining these ideas with those of the present paper, one would arrive at a description with covariant matrix fields $V^\mu(y^\nu)$ for all spacetime directions depending on spacetime rather than worldvolume coordinates. Amusingly, the description would look the same regardless of the dimension of the D-branes. The difference would be that for the lower dimensional branes, we would have a larger group of gauge symmetries, and therefore fewer physical degrees of freedom. It will be interesting to see whether these ideas can be realized more concretely.

\section{Discussion}\label{sec:discussion}

In this paper, we have seen that given a consistent transformation law for the matrix coordinates of D-particles, we can construct a covariant object $V^i_y$ transforming as a vector field under diffeomorphisms. Using this covariant object, we have described the most general actions consistent with general covariance. This allows us to write down the most general ``covariantization'' of any known leading order term (including leading order couplings to bulk fields other than the metric) thereby placing an infinite series of constraints on terms at higher orders.

By describing our branes using $V_y$ rather than $X$, we have in a sense promoted a worldvolume field to a spacetime field satisfying some constraints. Since it can be rather misleading to think of the worldvolume of a general noncommuting D-particle configuration as being pointlike (consider for example, the fuzzy sphere), it is appealing to have a description that does away with all reference to the worldvolume coordinates.

The construction presented here assumes that for the spaces under consideration there exists a unique spatial geodesic between any two points. Otherwise, the vector $V_y$ is well defined only in some neighborhood of the branes, even in the abelian case. The breakdown of our description for more general spaces is to be expected. For if we have two branes sitting at points connected by more than one geodesic, there will be more than one stable string configuration connecting the two branes, and therefore the number of fluctuating degrees of freedom will be larger than those described by a single matrix. Of course, this is very well known for the case of D-branes on tori, where the infinite number of geodesics connecting any two points require us to use infinite matrices to describe the D-brane degrees of freedom \cite{taylor}.   

A significant open problem is to prove that a consistent
transformation rule exists, or equivalently to show that our
constraints (\ref{covtens}) have a solution to all orders (we have
shown this to order $V^4$). Assuming that a solution exists, it is also necessary to understand whether the result for the transformation rule is unique, and if not, to interpret the arbitrary coefficients that appear. 

It will be interesting to extend our results to the case of higher-dimensional branes. For diffeomorphisms involving only the directions transverse to the branes, the story will be identical to that presented here. However, as discussed above, implementing invariance under diffeomorphisms that mix brane directions and transverse directions will likely require more serious work. 

Finally, it will be interesting to understand whether the class of
generally covariant actions presented here (or a more restricted class
that takes into account additional constraints) predicts any
interesting generic phenomena for D-branes on curved spaces, such as a
gravitational version of the dielectric effect \cite{sahakian1,sahakian2,grav_myers,hyakutake}.

\section*{Acknowledgments}

We would like to thank Miguel Costa, Michael Douglas, Carlos Herdeiro, Koenraad Schalm, Ehud Schreiber, and Washington Taylor for useful discussions and comments.  DB thanks the University of Porto for hospitality during part of this work. This work has been supported in part by the Natural Sciences and Engineering Council of Canada and by the Canada Research Chairs programme. DB and KF have been supported in part by postdoctoral fellowships from the Pacific Institute for the Mathematical Sciences. 

\appendix

\section{Conventions and useful formulae} \label{app:conventions}

In this paper, all metrics are spatial, with positive signature. The
Christoffel symbol is defined by
\[
\Gamma^i_{jk} = {1 \over 2} g^{il} (\partial_j g_{lk} + \partial_k g_{lj} - \partial_l g_{jk})
\]
and, under $y^i \rightarrow F^i(y)$, it transforms as
\be
\tilde{\Gamma}^i_{jk} = \partial_l F^i (\partial F)^{-1\,m}_j
(\partial F)^{-1\,n}_k \Gamma^l_{mn} - (\partial F)^{-1\,m}_j
\partial_m \partial_p F^i (\partial F)^{-1\,p}_k \; .
\label{eqn:chr_xfm}
\ee
In terms of this, the solution to the geodesic equation
\[
\ddot{x}^i + \Gamma^i_{jk} \dot{x}^j \dot{x}^k = 0
\]
with $x^i(0) = y^i$ and $\dot{x}^i(0) = v^i$ is given by
\be
\label{vsoln}
x^i = y^i + v^i - \sum_{n=2}^\infty \Gamma^i_{j_1 \cdots j_n} v^{j_1} \cdots v^{j_n}
\ee
where we define the $\Gamma$s recursively by
\[
\Gamma^i_{j_1 \cdots j_n} = \nabla_{(j_1} \Gamma^i_{j_2 \cdots j_n)} -
(n-1) \Gamma^i_{k (j_1} \Gamma^k_{j_2 \cdots j_n)} \; .
\]
Note that we have defined the $\Gamma$s all to be completely
symmetrized, and that an arbitrary local expression built from the metric and its derivatives may be written in terms of these quantities together with tensors built from the metric. 

Our conventions for the Riemann tensor are
\[
R^i_{jkl} = \partial_k \Gamma^i_{jl} - \partial_l \Gamma^i_{jk} + \Gamma^i_{mk} \Gamma^m_{jl} - \Gamma^i_{ml} \Gamma^m_{jk} \; .
\]

\section{Solving the constraints for $V_y$} \label{app:constraints}

We would like to determine the most general expression for $V_y(X,g)$ satisfying the constraints (\ref{covtens}) and consistent with (\ref{vseries}) in the abelian case. Defining $\Delta^i_y = X^i - y^i$, we begin with the definition
\be
\label{dser}
\Delta^i_y = V^i_y + \sum_{n \ge 2} \Delta^i_{j_1 \cdots j_n} V^{j_1}_y \cdots V^{j_n}_y \; .
\ee
To satisfy the linearity constraint (\ref{linear}) on the transformation law, the coefficients $\Delta^i_{j_1 \cdots j_n}$ must take the form (\ref{coeffconstr}). Further, we may assume that all the pure tensor terms vanish, since we can always consider redefinitions of the type (\ref{vredef}) at the end to recover the most general possibility. Thus, we may write
\be
\label{coeffconstr2}
\Delta^i_{j_1 \cdots j_n} = -{1 \over n!} \Gamma^i_{j_1 \cdots j_n} + \sum \Gamma^i_{j_1 \cdots j_m k_{m+1} \cdots k_{l}} (T_{\{nlm\}})^{ k_{m+1} \cdots k_{l}}_{j_{m+1} \cdots j_{n}}
\ee
where the first term is chosen for agreement with the abelian expression (\ref{vseries}). Here the $T$s are arbitrary tensors, but we assume that all coefficients are dimensionless, so that the numbers ${\cal N}_R$ and ${\cal N}_\nabla$ of curvatures and covariant derivatives in $T_{nlm}$ satisfy $2{\cal N}_R + {\cal N}_\nabla = n-l$ (more generally, we could have higher dimension corrections involving $\alpha'$ or some other dimensionful coefficient). Further, we may assume that the expression (\ref{dser}) contains no terms with $V$s contracted explicitly by the metric. For dimensional reasons, any such terms will have Riemann tensors with self contractions (e.g. Ricci tensors) or terms with lower indices on Riemann tensors contracted by explicit metrics (e.g. $R^{ijkl} R_{ijkl}$). The presence of these more general terms in the relation between $V$ and $X$ will lead only to the same types of terms on the right hand side of (\ref{covtens}), and conversely, such terms on the right hand side of (\ref{covtens}) may arise only from these sorts of terms in the relation between $V$ and $X$. Thus, given any $V_y(X,g)$ whose covariant derivative is a tensor $T$, we can keep only the terms in $V_y(X,g)$ satisfying our restriction, and the covariant derivative will be the tensor obtained from $T$ by dropping all terms not satisfying our requirement. 

Starting from (\ref{coeffconstr2}), we may invert the power series (\ref{dser}) to obtain
\be
\label{vser}
V^i_y = \Delta^i + \sum_{n \ge 2} V^i_{j_1 \cdots j_n} \Delta^{j_1} \cdots \Delta^{j_n} 
\ee
where the $V$ coefficients are determined in terms of the $\Delta$s by
\be
\label{rel2}
\sum_{n \le l} \Delta^i_{a_1 \cdots a_n}  \{ V^{a_1} \cdots V^{a_n} \}_{j_1 \cdots j_l} = 0 \qquad \qquad (l \ge 2) \; .
\ee
Here, the indices $\{ j_1, \cdots, j_l \}$ are to be distributed (in order) in all possible ways over the $V$s in curly brackets.

We may now constrain the coefficients by our requirement (\ref{covtens}) that the covariant derivative of $V_y$ should be a tensor. Thus, we demand that
\[
\nabla_i V^j = -T^j_{i; k_1 \cdots k_n} V^{k_1} \cdots V^{k_n} 
\]
where the coefficients $T$ must be tensors built from the metric (not related to those in (\ref{coeffconstr2})). Plugging in our expansion for $V$, this implies that
\be
\label{rel3}
V^i_{jk} + V^i_{kj} = \Gamma^i_{jk} 
\ee
and that for $l \ge 2$ we must have
\be
\label{rel4}
S^i_{j ; k_1 \cdots k_l} = T^i_{j; k_1 \cdots k_l} + \partial_j V^i_{k_1 \cdots k_l} + \sum_{n<l} (\partial_j V^i_{a_1 \cdots a_n}- S^i_{j ; a_1 \cdots a_n}) \{ \Delta^{a_1} \cdots \Delta^{a_n} \}_{j_1 \cdots j_l}
\ee
where we have defined
\[
S^i_{j ; k_1 \cdots k_l} = V^i_{j k_1 \cdots k_l} + V^i_{ k_1 j \cdots k_l} + \cdots + V^i_{k_1 \cdots k_l j} \; .
\]
To solve the constraints, we write the most general expressions for the $T_{\{nlm\}}$s in (\ref{coeffconstr2}), use (\ref{rel2}) to write the $V$s in terms of these expressions, and then use (\ref{rel3}) and (\ref{rel4}) to constrain the coefficients. Note that in applying (\ref{rel4}), we also must allow $T^i_{j; k_1 \cdots k_l}$ to be an arbitrary tensor.

\subsubsection*{The transformation law and covariant vector field at fourth order}

Up to fourth order in $V$, we find that the most general possible solution of the explicit constraints we have just outlined (with all tensor terms vanishing) is 
\bea
\Delta^i &=& V_y^i - {1 \over 2} \Gamma^i_{jk} V^j_y V^k_y - {1 \over
3!} \Gamma^i_{jkl} V^j_y V^k_y V^l_y - {1 \over 4!} \Gamma^i_{jklm} V^j_y V^k_y V^l_y V^m_y \cr 
&& \qquad \qquad \qquad + {1 \over 18} \Gamma^i_{jn} R^n_{klm} \left\{
V^m[V^{(k},[V^{l)},V^j]] + C^{jklm}(V) \right\} + {\cal O}(V^5) \; .
\label{fourthorder}
\eea
Here, the expression $C^{ijkl}(X)$ indicates any complete commutator expression, 
\[
\partial_{y^m} C^{ijkl}(X+y) = 0 \; .
\]
There are six independent terms of this type,
\beas
C^{ijkl}(X) &=& a[X^i,X^j][X^k,X^l] + b[X^i,X^k][X^j,X^l] + c[X^i,[X^j,[X^k,X^l]]] \cr
&&+ d [X^j,[X^i,[X^k,X^l]]] + e [X^k,[X^i,[X^j,X^l]]] + f
[X^k,[X^j,[X^i,X^l]]] \; .
\eeas
With this definition, the covariant derivative of $V$ is a tensor given by
\be
\label{covcan}
\nabla_i V^j = -\delta_i^j -[{1 \over 6}  R^j_{kli}  + {1 \over 6}
R^j_{lki}] V^k V^l - {1 \over 12} \nabla_{(k}R^j_{lm)i} V^k V^l V^m +
{\cal O}(V^4) \; .
\ee
Note that the arbitrary coefficients in (\ref{fourthorder}) do not affect (\ref{covcan}) until the next order.

From the relation between $X$ and $V_y$, we find the following
transformation law for $X$ up to fourth order in $X$ (all quantities are evaluated at $y=0$):
\bea
\label{fourth}
\Phi^i(F,X,g) &=& F^i \cr
&& + \partial_j F^i X^j \cr
&& + {1 \over 2} \partial_j \partial_k F^i X^j X^k \cr
&& + {1 \over 3!} \partial_j \partial_k \partial_l F^i X^j X^k X^l  - {1 \over 12} \partial_m \partial_k F^i \; \Gamma^m_{jl} [X^j,[X^k,X^l]]\cr
&& + {1 \over 4!} \partial_j \partial_k \partial_l \partial_m F^i X^j X^k X^l X^m \cr
&& + {1 \over 24} \partial_j \partial_k \partial_n F^i \; \Gamma^n_{lm} (X^l X^m X^j X^k + X^j X^l X^m X^k + X^j X^k X^l X^m \cr 
&& \qquad - X^j X^l X^k X^m - X^l X^j X^m X^k - X^l X^j X^k X^m) \cr
&& + {1 \over 24} \partial_j \partial_n F^i \; \Gamma^n_{klm} (X^k X^l X^m X^j - X^k X^l X^j X^m - X^k X^j X^l X^m + X^j X^k X^l X^m) \cr
&& + {1 \over 24} \partial_n \partial_p F^i \; \Gamma^n_{jk} \Gamma^p_{lm}(X^j X^l X^k X^m - X^j X^l X^m X^k)  \cr
&& + {1 \over 24} \partial_j \partial_n F^i \; \Gamma^n_{kp} \Gamma^p_{lm} (X^j X^l X^m X^k  + X^j X^k X^l X^m - 2 X^l X^m X^j X^k \cr 
&& \qquad + X^k X^l X^m X^j + X^l X^m X^k X^j - 2 X^k X^j X^l X^m) \cr
&& - {1 \over 18} \partial_j \partial_n F^i \; R^n_{klm} (X^m [X^{(k},
[X^{l)}, X^j]] + C^{jklm}(X))  + {\cal O}(X^5) \; .
\eea
Thus, all of the arbitrary coefficients in $V$ appear in the transformation law, as we should expect, since we have already fixed any ambiguity associated with transformation of the type (\ref{vredef}) by assuming all pure tensor terms in $\Delta(V)$ vanish. Various possible interpretations for these arbitrary coefficients are suggested in section 3.

\subsubsection*{Useful expressions for expanding actions about a point}

In writing out the component forms of the actions using (\ref{actexp}) we require expressions for the Taylor series coefficients 
\[
W^i_{i_1 \cdots i_n} = {1 \over n!} \partial_{i_1} \cdots \partial_{i_n} (V^i_y + y^i)|_{y=0}
\]
in the expansion of $V_y + y$ in powers of $y$. From (\ref{fourthorder}), we find
\begin{eqnarray}
W^i &=& X^i + {1 \over 2} \Gamma^i_{jk} X^j X^k + {1 \over 6}
(\Gamma^i_{jkl} + 3 \Gamma^i_{m(j}\Gamma^m_{kl)})X^j X^k X^l + {1
\over 12} \Gamma^i_{mj} \Gamma^m_{kl}[X^k,[X^l,X^j]] + {\cal O}(X^4) \cr
W^i_j &=& - \Gamma^i_{jk} X^k - {1 \over 2} (\Gamma^i_{jkl} + 3
\Gamma^i_{m(j}\Gamma^m_{kl)}) X^k X^l + {1 \over 2} \partial_j
\Gamma^i_{kl} X^k X^l + {\cal O}(X^3) \cr
W^i_{jk} &=& {1 \over 2} \Gamma^i_{jk} + {1 \over 2}(\Gamma^i_{jkl} +
3 \Gamma^i_{m(j} \Gamma^m_{kl)}) X^l - \partial_{(j} \Gamma^i_{k)l}
X^l + {\cal O}(X^2) \cr
W^i_{jkl} &=& - {1 \over 6} \Gamma^i_{jkl} - {1 \over 2}
\Gamma^i_{n(j} \Gamma^n_{kl)} + {1 \over 2} \partial_{(j}
\Gamma^i_{kl)} + {\cal O}(X) \; .
\end{eqnarray}
As discussed in section \ref{sec:actions}, we can replace $X$ by $V$ in any expanded action to obtain a scalar position-independent expression depending only on $V$ and tensors built from the metric. This may be obtained most simply by working in Riemann normal coordinates, so it is also useful to have expressions for $W$s in Riemann normal coordinates. Going to this system, we have for example, 
\beas
\partial_j \Gamma^i_{kl} &\to& -{2 \over 3} R^i_{(kl)j} \cr
\partial_j \Gamma^i_{klm} &\to& - {1 \over 2} \nabla_{(k} R^i_{lm)j} \cr
\partial_j \partial_k \Gamma^i_{lm} &\to& -{5 \over 6} \ R^i_{(lm)(j;k)} + {1 \over 6} R^i_{(jk)(l;m)}
\eeas
so that we obtain
\begin{eqnarray}
W^i &=& X^i + {\cal O}(X^4)\cr
W^i_{j} &=& -{1 \over 3} R^i_{(kl)j} X^k X^l - {1 \over 12}
\nabla_{(k}R^i_{lm)j} X^k X^l X^m + {\cal O}(X^4) \cr
W^i_{jk} &=& -{1 \over 3} R^i_{(jk)l} X^l - {1 \over 24} \nabla_{l}
R^i_{(jk)m} X^{(l} X^{m)} - {1 \over 8} \nabla_{(j} R^i_{|lm|k)} X^{(l}
X^{m)} + {\cal O}(X^3) \cr
W^i_{jkl} &=& -{1 \over 6} \nabla_{(j} R^i_{kl)m} X^m + {\cal O}(X^2) \cr
W^i_{jklm} &=& {\cal O} (\nabla \nabla R X) \; .
\end{eqnarray}

\section{A technical result about commutator expressions}\label{app:technical}

In this appendix, we argue that for any expression ${\cal L} = A_{i_1 \cdots i_{n}} X^{i_1} \cdots X^{i_n}$ with more than three $X$s such that $\tr({\cal L})$ satisfies (\ref{cond1}), that is
\be
\label{ccc1}
\partial_{X^i} \tr({\cal L}) = 0 \; ,
\ee
there is another expression ${\cal L}'$, equivalent under the trace (i.e. $ \tr({\cal L}') = \tr({\cal L})$) satisfying (\ref{cond2}),
\[
\partial_{X^i} {\cal L}' = 0 \; .
\]
For simplicity, we discuss terms without $\dot{X}$s, but our argument can easily be extended to include them since they trivially satisfy $\partial_{X^i} \dot{X} = 0$.

Note first that any symmetry properties of the coefficients $A$ can be transferred directly to the product of $X$s by replacing the ordering shown with an average over all orderings that give the same result when contracted with $A$. Writing the resulting sum over permutations as  
\[
{\cal O}^{i_1 \cdots i_n} = \sum_\sigma a_{\sigma}  X^{i_{\sigma(1)}} \cdots X^{i_{\sigma(n)}} \; .
\]
It now follows that (\ref{ccc1}) will be satisfied if and only if 
\be
\label{cc1}
\partial_{X^i} \tr({\cal O}^{i_1 \cdots i_n}) = 0 \; , 
\ee
since any cancellations that resulted from symmetries of $A$ will still occur here. Now, by rearranging terms in the trace, we can write all terms such that $X^{i_1}$ appears first,
\be
\label{equiv}
\tr({\cal O}^{i_1 \cdots i_n}) = \tr(X^{i_1} {\cal O}^{i_2 \cdots
i_n}_1) \; .
\ee
Then (\ref{cc1}) implies that
\[
0 = y^i \partial_{X^i} \tr({\cal O}^{i_1 \cdots i_n}) = y^{i_1} \tr({\cal O}_1^{i_2 \cdots i_n}) + \sum_m y^{i_m} \tr(X^{i_1} \partial_{X^{i_m}} {\cal O}_1^{i_2 \cdots i_n}) \; .
\]
In order that the right side should vanish, the expressions involving $y^{i_m}$ must vanish independently for each $m$, therefore 
\[
\tr({\cal O}_1) = 0 \qquad \qquad \partial_{X^{i}} {\cal O}_1 = 0 \; .
\] 
The second condition implies that ${\cal O}_1$ is a sum of expressions
built from products of nested commutators.\footnote{Equivalently,
${\cal O}_1$ must be the dimensional reduction to 0+0 dimensions of
some gauge covariant expression.}  Since the trace of ${\cal O}_1$ must also vanish, we should have 
\[
{\cal O}_1 =  \sum_{\alpha} [{\cal A}_\alpha ,{\cal B}_\alpha] 
\]
where ${\cal A}_\alpha$ and ${\cal B}_\alpha$ each must take the form of either a single matrix $X$ or an expression built completely from products of nested commutators. Since we assumed that ${\cal O}^{i_1 \cdots i_n}$ contained at least four $X$s, at least one of ${\cal A}_\alpha$ and ${\cal B}_\alpha$ must contain more than a single $X$ for each $\alpha$ (we may assume, without loss of generality that it is ${\cal B}_\alpha$).  Then, inserting this expression into (\ref{equiv}) and rearranging the commutator, we find
\[
\tr({\cal O}) = \tr({\cal O}')
\]
where
\be
\label{equi}
{\cal O}' = \sum_\alpha [X^{i_1}, {\cal A}_{\alpha}] {\cal B}_{\alpha} \; .
\ee
Further, it is clear that $\partial_{X^i} {\cal O}'=0$ since ${\cal O}'$ is a sum of complete commutator expressions. Finally, if we define
\[
{\cal L}' = A_{i_1 \cdots i_n} ({\cal O}')^{i_1 \cdots i_n}
\]
then $\tr({\cal L}') = \tr({\cal L})$ and $\partial_{X^i}{\cal L}' =0$
as desired.  

Note that for terms at order $X^3$, both ${\cal A}$ and ${\cal B}$ must be single $X$s, and as a result our assertion fails.

\section{Promoting the world-volume gauge symmetry to a local spacetime symmetry}\label{app:promote}

Since we have promoted the matrix $X$ to a covariant matrix field $V^i_y$, it is interesting to ask whether we can also promote the gauge symmetry $X^i(t) \to U(t) X^i(t) U^{-1}(t)$ to a symmetry that is local on the transverse space, $V^i_y \to U(y) V^i_y U^{-1}(y)$. As it stands, this generalization leads to an ill-defined transformation rule for $X$ unless $U(y)$ is constant. However, if we introduce in addition to $V_y$ a flat gauge connection $A_i(y)$ on the transverse space, then we can define a correspondence
\[
(V^i_y, A_i(y)) \to (\hat{V}_y^i, 0) \to X^i
\]
which associates a well-defined matrix $X$ with any pair $(V,A_{flat})$. Here, the first step is the local gauge transformation $V \to U V U^{-1}, A_i \to U A_i U^{-1} - i \partial_i U U^{-1}$ that brings $A$ to zero, while the second step is our previous correspondence between $X$ and $V$. With this correspondence, any local gauge transformation acting on the left side will reduce to a well defined transformation on $X$. With this more general description, the constraint (\ref{covtens}) should be modified to include the gauge connection in the covariant derivative on the left side. It remains to be seen whether this more general description will turn out to be useful.

\end{document}